\pdfoutput=1
\documentclass[doublecol]{epl2}
\usepackage[english]{babel}
\usepackage[latin1]{inputenc}
\usepackage{amsmath,bbm}
\usepackage{graphicx}
\usepackage{amsfonts}

\newcommand{\bfe}{{\mathbf{e}}}

\newcommand{\bfk}{{\mathbf{k}}}
\newcommand{\bfN}{{\mathbf{N}}}
\newcommand{\bfq}{{\mathbf{q}}}
\newcommand{\bfr}{{\mathbf{r}}}

\newcommand{\bfu}{{\mathbf{u}}}

\newcommand{\bfx}{{\mathbf{x}}}

\newcommand{\calA}{{\mathcal{A}}}
\newcommand{\calB}{{\mathcal{B}}}

\newcommand{\calG}{{\mathcal{G}}}

\newcommand{\calQ}{{\mathcal{Q}}}

\newcommand{\hatD}{\hat{D}}

\newcommand{\hatu}{\hat{u}}
\newcommand{\hatbfu}{\hat{\bfu}}
\newcommand{\hatcalG}{\hat{\calG}}
\newcommand{\hatcalQ}{\hat{\calQ}}

\newcommand{\barx}{\bar{x}}
\newcommand{\barbfx}{\bar{\bfx}}

\newcommand{\tilmu}{\tilde{\mu}}
\newcommand{\tillambda}{\tilde{\lambda}}

\newcommand{\Reals}{\mathbbm{R}}

\begin{document}

\title{Elastic fluctuations as observed in a confocal slice}
\author{Claire A. Lemarchand, A. C. Maggs, Michael Schindler}
\institute{Laboratoire PCT, UMR Gulliver CNRS-ESPCI 7083, 10 rue Vauquelin, 75231 Paris Cedex 05}

\date{\today}

\abstract{
  Recent confocal experiments on colloidal solids motivate a fuller study of the projection
  of three-dimensional fluctuations onto a two-dimensional confocal
  slice. We show that the effective theory of a projected crystal
  displays several exceptional features, such as non-standard
  exponents in the dispersion relations. We provide
  analytic expressions for the effective two-dimensional elastic
  properties which allow one to work back from sliced experimental
  observations to three-dimensional elastic constants.
}
\maketitle

Optical techniques, including scattering and microscopy, have long
been used to extract detailed static and dynamic information from soft
condensed matter systems. They are in many ways complementary --
scattering being most suitable for examining the fluctuations in
Fourier space~\cite{chaikin}, giving information on the mode structure
for uniform systems; microscopy gives the best account of the real
space structure of a medium~\cite{weeks2} and is particularly useful
in the study of heterogeneous material properties.  Recently several
experimental groups have studied the properties of colloidal crystals
using video and confocal microscopy at
interfaces~\cite{maret,maret2,andrea} and in full three-dimensional
samples \cite{maret3}.  Using computer analysis one combines the
advantages of scattering and direct observation: One can observe a
carefully chosen part of an experimental system and then study the
mode structure of thermally excited fluctuations.  It is particularly
interesting to work deeply within a three dimensional
sample~\cite{antinaPRL,kurchan,ghosh,science} because one can be sure
that surface perturbations to the properties are weak.

However, the measurement of fluctuations rather than just the mean
positions of the particles is technically difficult: The scan must be
fast in order to freeze the particle motion during the acquisition of
a frame. Because of the constraints several groups have chosen to perform
observations on single confocal slices, rather than scanning the full
three dimensional volume.  The microscope resolves the motion of
the colloidal particles within a single plane of the crystal
structure--typically the plane $(1,1,1)$. The sample is filmed for several
minutes and the  matrix of correlations is generated within the slice. Many
thousands of frames are required in order to generate good fluctuation
statistics. Experimentalists have studied weakly \cite{science} or
strongly disordered \cite{kurchan,danielPRL} materials with the hope of better
understanding glassy dynamics, and characterizing the disorder via the
projected fluctuation properties including the spectrum
\cite{kurchan}, the eigenvectors \cite{antinaPRL} and the effective
dispersion relations \cite{science}. We show here that even in the
case of {\sl ordered} elastic materials a number of interesting
features appear. Exceptional behaviour in disordered materials should
thus be defined with respect to the conclusions we present
here. Observation of non-standard exponents for correlations within
the slice can not be interpreted immediately as evidence of exotic or glassy behaviour.


Our paper aims to calculate the effective theory which best describes
the fluctuations of such a two-dimensional slice of a larger
three-dimensional sample, in order to be able to easily work back from
the observed two-dimensional correlations to three-dimensional
material constants. The link between these elastic constants is given in Eqs.~\eqref{eq:dispersion}
below, which is the main result of the paper. The calculation of the projected properties will be based on
the following principles: At large length scales a colloidal crystal
is described by an effective elastic
theory~\cite{chaikin,maret2,ChaLub95}; such an elastic theory leads to
Gaussian fluctuations. Gaussian systems are rather special in that
they allow one to exactly trace out degrees of freedom leading to a
new effective theory which is also Gaussian in nature. This effective
theory requires, however, effective {\sl renormalized} couplings.
Arguments due to Peierls~\cite{peierls} immediately show that the
final theoretical description of an elastic slice must be unusual: A
two-dimensional elastic solid has diverging fluctuations in the
positions of individual particles, whereas the slice of the
three-dimensional solid must have bounded fluctuations. The effective
theory describing the slice can not be a simple variation on standard
elasticity theory, we understand at once that anomalous dispersion
relations are to be expected.

Indeed we will show that while standard elasticity gives rise to a
scaling of the elastic energy in~$q^2$ as a function of the
wavevector, $\bfq$, the projected effective theory for the confocal
slice is characterized by an effective dispersion relation in
$|\bfq|$. We give explicit analytic expressions for the prefactors in
the dispersion law as a function of the three-dimensional elastic
moduli, see Eqs.~\eqref{eq:dispersion} below, {\it allowing one to deduce the
three-dimensional properties from measurements on the slice}. The
procedure described in this letter, albeit applied to elasticity, is
sufficiently general that similar scalings in~$|\bfq|$ should be
observed in very different projected systems, such as in Stokesian
hydrodynamics.

We begin by linking the fluctuations in an elastic solid to the static
elastic Green function of the medium. We then show how this Green
function can be projected into a single layer, to produce an effective
theory for the observed slice. We have performed extensive numerical
simulations, which we compare with the analytic theory.

Let us now consider a three-dimensional cubic crystal. Under small deformations
the system is characterized by the displacement vector $u_i$ and the
symmetric homogeneous tensor of displacement
gradients~$u_{ij}$~\cite{landau}. The elastic energy is then written
as a quadratic form in~$u_{ij}$ which respects the cubic symmetry of
the crystal. This quadratic form is related to the elastic
matrix~\cite{Wallace70}
\begin{equation}
  \label{eq:christoffel}
  \hatD_{ik}(\bfk) = \Bigl[\lambda\delta_{ij}\delta_{kl}
  + \mu(\delta_{ik}\delta_{jl} + \delta_{il}\delta_{jk})
  + \nu S_{ijkl}\Bigr] k_j k_l,
\end{equation}
with Lam\'e constants $\lambda$, $\mu$ and anisotropy
$\nu$.\footnote{\label{fn1}Note that if the reference configuration of
the crystal is under external stress, this stress appears explicitly
in the elastic tensor~\cite{Wallace70}. A hard-sphere crystal is
always under external pressure to be mechanically stable. The elastic
constants in Eq.~\eqref{eq:christoffel} implicitly contain this
pressure correction.} The hat denotes a Fourier transform and
summation over repeated indices is assumed throughout the paper. The
tensor $S=\sum_{p=1}^3 \bfe^{(p)} \bfe^{(p)} \bfe^{(p)} \bfe^{(p)}$,
with ${\bfe}^p$ unit vectors parallel to the cubic axes of the
crystal. The Green function of the static elastic problem is then the
inverse of the elastic tensor,
\begin{equation}
  \label{eq:inverse}
  \hatD_{ij}(\bfk) \hatcalG_{jk}(\bfk) = \delta_{ij}.
\end{equation}%
One expresses the free energy in terms of the displacement field
\begin{equation}%
  F[\hatbfu] = \frac{1}{2} \sum_{\bfk} \hatu_i(\bfk)\hatD_{ij}(\bfk)\hatu_j(\bf-k)\text{.}
\end{equation}%
If the crystal is studied at a finite inverse temperature $\beta$,
this implies that the correlation in the fluctuation amplitudes is
given by
\begin{equation}
  \label{eq:canonical_k}
  \bigl\langle \hatu_i(\bfk) \hatu_j(-\bfk)\bigr\rangle
  = \frac{1}{Z} \int_{\Reals^3} d\hatbfu\: e^{-\beta F[\hatbfu]} \hatu_i \hatu_j
  = \beta^{-1}\hatcalG_{ij}(\bfk).
\end{equation}
For each wavevector $\bfk$, $\hatD$ is a $3\times3$~matrix
with eigenvalues $d_i({\bf k})$ where the subscript $i$~indicates a
polarization state. Following a convention usual in the experimental
literature~\cite{maret2,antinaPRL}, we define the auxiliary
variable~$\omega^2_i(\bfk) = d_i({\bf k})$.\footnote{Notice that
$\omega$ is not the angular velocity of a wave.}

Instead of the full crystal, we now consider a crystal layer observed
in a confocal microscope.  In the following, $\calQ_{\alpha\beta}$~is
the Green function reduced to two dimensions,
$\alpha,\beta\in\{1,2\}$, and $\bfx,\bfq\in\Reals^2$~are direct and
reciprocal vectors in reduced space, whereas their three-dimensional
counterparts are denoted $\bfr,\bfk\in\Reals^3$.  Of course,
neglecting the third dimension does not change the correlations within
the layer; the real-space Green functions of the projected and of the
full problem are the same.

From the Green function in the reduced space we then perform an
inverse, two-dimensional, transform to find the effective dispersion
relation for the observed slice. We wish to describe the fluctuations
in the two-dimensional plane using a closed theory, calculating the
two-dimensional equivalent of the matrix~$\hatD$. The calculational
route that we will follow is
\begin{multline}%
  \label{succession}%
  \hatD_{ij}(\bfk) \longrightarrow \hatcalG_{ij}(\bfk)
  \overset{\mathcal{F}^{-1}_3}{\longrightarrow} \calG_{ij}(\bfr) \\
  \longrightarrow \calQ_{\alpha\beta}(\bfx)
  \overset{\mathcal{F}_2}{\longrightarrow}
  \hatcalQ_{\alpha\beta}(\bfq) \longrightarrow
  \hatD_{\alpha\beta}(\bfq),
\end{multline}%
where $\mathcal{F}_l$ is a $l$-dimensional Fourier transform.

The Green function~$\hatcalG_{ij}(\bfk)$ in three-dimensional
reciprocal space follows from the inversion of the elastic
matrix~$\hatD_{ij}$ in Eq.~\eqref{eq:inverse}. The result will have
the following tensorial form,
\begin{equation}%
  \label{eq:Grecip}
  \hatcalG_{ij}(\bfk) = \frac{1}{k^2}\Bigl[
  A'\,\delta_{ij}
  + A''\,\frac{k_ik_j}{k^2}
  + A'''\calA_{ij}(\bfk)\Bigr],
\end{equation}%
where the prefactors~$A',A'',A'''$ are cubic scalars and are not
required to be isotropic; they may depend on the orientation
of~$\bfk$. The first two terms in the brackets are the nearly
isotropic part, while the third term~$\calA_{ij}$ comprises further
anisotropic properties. An explicit form of the latter can be obtained
either by direct inversion of~$\hatD_{ij}$ using the Sherman--Morrison
formula or in a more elaborate way using bases for the space of all
cubic tensors. The scalar prefactors are determined from the linear
system of three equations which are obtained after multiplication of
Eq.~\eqref{eq:inverse} with $\delta_{ij}$, $k_ik_j/k^2$, and
$\calA_{ij}$, respectively. The Fourier transform has a similar
tensorial structure,
\begin{equation}%
  \label{eq:Gdirect}
  \calG_{ij}(\bfr) = \frac{1}{4\pi r}\Bigl[
  B'\,\delta_{ij}
  + B''\,\frac{r_ir_j}{r^2}
  + B'''\calB_{ij}(\bfr)\Bigr].
\end{equation}%
Again, the cubic scalars~$B',B'',B'''$ depend on the orientation
of~$\bfk$ and are obtained as the solution of a linear system of
equations.

The reduction to the two-dimensional Green
function~$\calQ_{\alpha\beta}(\bfx)$ can now be performed in real
space. We firstly recognize that the result will be nearly isotropic,
since the crystal plane we project on is a hexagonal
lattice~\cite{landau}. The Green function comprises two tensorial
parts\footnote{If the external stress mentioned in note~\ref{fn1} were
anisotropic, a third tensorial term would be allowed by symmetry.},
\begin{equation}%
  \label{eq:Qdirect}
  \calQ_{\alpha\beta}(\bfx) = \frac{1}{4\pi x}\Bigl[
  C'(\bfx)\delta_{\alpha\beta}
  + C''(\bfx)\frac{x_\alpha x_\beta}{x^2} \Bigr].
\end{equation}%
where again the hexagonal scalar prefactors may depend on the
orientation of~$\bfx$. They are determined by projection of the
three-dimensional Green function onto the two-dimensional subspace.
For doing this, we choose an orthonormal basis
$(\bfr^{(1)},\bfr^{(2)},\bfN)$, aligned such that
$\bfN=(1,1,1)/\sqrt{3}$~is orthogonal to the plane we project on. We
now interpret $\bfx$~as a three-dimensional vector, denoted by an
overbar, $\bar\bfx = \sum_{\alpha=1}^2 x_\alpha\bfr^{(\alpha)}$. The
two-dimensional identity tensor~$\delta_{\alpha\beta}$ then becomes
$[\delta_{ij} - N_iN_j]$ in three dimensions. Reduction of
Eqs.~\eqref{eq:Qdirect} and \eqref{eq:Gdirect} by
$\delta_{\alpha\beta}$ and $x_\alpha x_\beta/x^2$ and by their
three-dimensional counterparts, allows to determine the prefactors in
Eq.~\eqref{eq:Qdirect},
\begin{subequations}%
  \begin{align}%
    \frac{2C'(\bfx) + C''(\bfx)}{4\pi x} &= \bigl[\delta_{ij} - N_iN_j\bigr] \calG_{ij}(\barbfx), \\
    \frac{C'(\bfx) + C''(\bfx)}{4\pi x} &= \frac{\barx_i\barx_j}{x^2}
    \calG_{ij}(\barbfx).
  \end{align}%
\end{subequations}%
In the same way as in three dimensions, we obtain the reduced Green
function in reciprocal space using the ansatz
\begin{equation}%
  \label{eq:Qrecip}
  \hatcalQ_{\alpha\beta}(\bfq) = \frac{1}{2q}\Bigl[
  E'(\bfq)\delta_{\alpha\beta}
  + E''(\bfq)\frac{q_\alpha q_\beta}{q^2} \Bigr].
\end{equation}%

During the manipulations from Eq.~\eqref{eq:Grecip} to
Eq.~\eqref{eq:Qrecip} the scalar prefactors inherited the nontrivial
vector-dependence from each other. It is important to notice that this
dependence is restricted to the \emph{orientation} of the vectors,
since we regard only the long-wavelength limit, in which the elastic
moduli in Eq.~\eqref{eq:christoffel} are constants. The dependence on
the \emph{norm} of the vector is written out explicitly in
Eqs.~\eqref{eq:Grecip}--\eqref{eq:Qrecip}. In particular, the
two-dimensional Fourier transform of~$\calQ(\bfx)\sim 1/x$ led to the
scaling $\hatcalQ(\bfq)\sim 1/q$.

Using only symmetry arguments, the scalar prefactors in
Eq.~\eqref{eq:Qrecip} cannot be expected to be fully isotropic or even
constants. In the numerical simulation described below, we observe
however that in the limit of small~$q$ their angular dependence is
negligible. Unfortunately, the direct-space cubic Green
function~$\calG_{ij}(\bfr)$ cannot be calculated explicitly, except
for a few directions of higher symmetry~\cite{Morawiec94}. A more
practical way is to approximate the cubic Green function by an
appropriate isotropic one. Fedorov~\cite{Fedorov68,Norris06} provided
an optimal way to do this, based on slowness curves. He proposed the
effective isotropic moduli
\begin{equation}
  \label{eq:fedorov}
  \tillambda = \lambda + \frac{\nu}{5}, \quad\text{and}\quad
  \tilmu = \mu + \frac{\nu}{5},
\end{equation}
which give the optimal three-dimensional Green function
\begin{equation}%
  \label{eq:iso_green}
  \hatcalG_{ij}(\bfk) = \frac{1}{k^2}\Bigl[
  \frac{1}{\tilmu} \delta_{ij}
  -\frac{\tillambda+\tilmu}{\tilmu(\tillambda+2\tilmu)} \frac{k_i
    k_j}{k^2}\Bigr].
\end{equation}%
The scalar prefactors in this and in the other Green functions are
constants; $A'$~and $A''$ in terms of $\tilmu,\tillambda$ can be read
off Eq.~\eqref{eq:iso_green}; $A'''=B'''=0$ by isotropy; for the
others, we find $B'=C'=A'+A''/2$, $B''=C''=-A''/2$, $E'=A'$, and
$E''=A''/2$. The latter two give the isotropic projected Green
function from Eq.~\eqref{eq:Qrecip}, which we choose here to write in
its longitudinal/transverse form
\begin{equation}%
  \hatcalQ_{\alpha\beta}(\bfq) =
  \frac{1}{2\tilmu q} \Bigl(\delta_{\alpha\beta}{-}\frac{q_\alpha q_\beta}{q^2}\Bigr) +
  \frac{1}{q} \frac{\tillambda+3\tilmu}{4\tilmu(\tillambda{+}2\tilmu)}\frac{q_\alpha q_\beta}{q^2}.
\end{equation}%
The effective elastic matrix~$\hatD_{\alpha\beta}(\bfq)$ of the
projected two-dimensional slice, which is the inverse of this Green
function, thus predicts the following effective dispersion relations:
\begin{equation}
  \begin{aligned}
    \label{eq:dispersion}
    \omega_\perp^2 &= 2\tilmu\, q\quad\text{(transverse)}, \\
    \omega_\parallel^2 &=
    \frac{4\tilmu(\tillambda{+}2\tilmu)}{\tillambda+3\tilmu}\,
    q\quad\text{(longitudinal).}
  \end{aligned}
\end{equation}
These equations explain the anomalous (non-Debye) scaling of the
projected fluctuations, where the auxiliary variables~$\omega$
are proportional to~$\sqrt{q}$ at long wavelengths, rather than $q$  in standard elastic theory. The scaling is the
same for all branches. The prefactors are here given explicitly in terms
of isotropic approximations of the cubic elastic moduli for the
hexagonal $(1,1,1)$ plane in the face centered cubic crystal. For other 
modes and other orientations of the crystal explicit expressions are difficult to
obtain, but numerical methods can help \cite{SchMag11}. These simple
expressions can be used by the experimentalist to obtain quantitative
information on the three-dimensional elastic properties by studying
only the two-dimensional slice.

The distinction between the longitudinal and transverse branches has
already been noted in the experimental reports \cite{science, ghosh},
which however do not try to fit the experimental prefactors in order
to measure three-dimensional properties.
\begin{figure}[htb]%
  \begin{center}
    \includegraphics{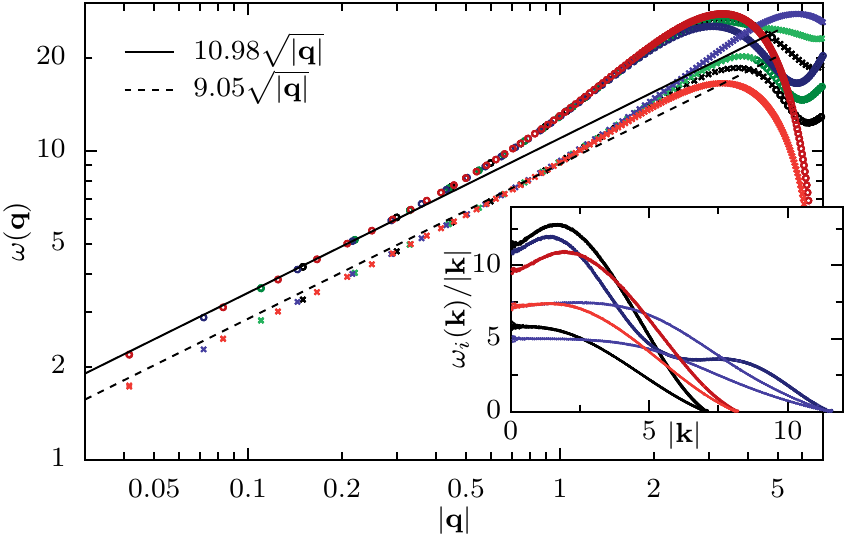}%
    \caption{Inset: Full, three-dimensional dispersion curves
    $\omega_i/k$ used to extract the three elastic constants of a
    cubic crystal. We plot the longitudinal and transverse modes in
    the directions $(1,0,0)$ (red), $(1,1,0)$ (blue) and $(1,1,1)$
    (black). The upper curves are longitudinal. For the orientations
    $(1,0,0)$ and $(1,1,1)$ the transverse modes are degenerate. Main curves:
    Dispersion relations in the confocal cut evaluated for the
    directions $(1,0)$, $(2,1)$, $(3,1)$, $(4,1)$ with respect to the
    hexagonal Bravais lattice. The lines are the analytic prediction
    for the long-wavelength limit, Eqs.~\eqref{eq:dispersion}
    and~\eqref{eq:fedorov}. $58,000$ hours of computer
    time.}%
    \label{fig:3D2D}%
  \end{center}
\end{figure}%
%
\medskip

To better understand the limits of our theoretical calculation we
performed a molecular dynamics simulation with $N=4,147,200$ particles
organized in a face-centred cubic crystal, with a volume fraction
$\phi=0.57$. We used event driven methods~\cite{rapaport} because of
their efficiency and also their long-time stability. The periodic
simulation box contained $N_1\times N_2\times N_3 = 160\times160\times162$ particles and had a
skew shape, aligned with the Bravais lattice. This choice allows easy
data analysis since the simulation box is also aligned with the slice
direction~$(1,1,1)$, and it avoids ambiguities in the definition of
the reciprocal vectors~$\bfk$.

In order to study the mode structure of fluctuations we calculated and
recorded the time average $\langle\hatu_i(\bfk)\hatu_j(-\bfk)\rangle$
in three dimensions and deduced the polarization eigenstates  by
diagonalizing the three dimensional matrix of correlations measured
for each $\bfk$. For the three-dimensional modes we find the expected
scaling $d_i({\bf k}) \sim k^2$ so that $\omega_i \sim k$. Thus when
we plot $\omega_i/k$ as a function of~$k$ (inset of
Fig.~\ref{fig:3D2D}) we can relate the small wavevector intersect to
the three-dimensional isothermal elastic constants in
Eq.(\ref{eq:christoffel}). The transverse and longitudinal dispersion
curves are then uniquely identified by their degeneracy\footnote{see
page~353 of Ref.~\cite{Wallace70}}, and they contain sufficient
information to extract the three independent elastic constants of a
cubic crystal~\cite{frenkel}. We find the numerical values
$\lambda,\mu,\nu=42.8,51.8,-53.8$ Units are chosen such that diameter,
mass of all particles, and $kT$ are unity.

We now repeat the analysis for the sliced fluctuations of the
crystal, $\langle\hatu_\alpha(-\bfq)\hatu_\beta(\bfq)\rangle$, using
the hexagonal $(1,1,1)$ planes to reproduce the experimental situation
of \cite{science,antinaPRL,thesis}, and extract the corresponding
eigenvalues and auxiliary variables, $\omega_i({\bf q})$, from the
resulting $2\times2$ matrices. The result is plotted in the main
figure of Fig.~\ref{fig:3D2D}. We see that two branches are important
at long wavelengths, and that as predicted in our analytic theory the
dispersion relation for the modes are of the form $\omega^2_i(\bfq)
\sim q$. Anomalous projected dispersion curves have already been
observed experimentally in both disordered and crystalline materials
\cite{science,thesis}.  Using the effective elastic constants of
Eqs.~\eqref{eq:fedorov} together with the dispersion law
eq.~\eqref{eq:dispersion}, we calculate the prefactors to this law and
plot the results as the lines in Figure \ref{fig:3D2D}. The
theoretical and measured curves agree to within $3\%$.
We interpret these discrepancies as being due to the anisotropy of the
crystal. We note that a face-centred cubic crystal with
nearest-neighbour interactions is predicted in linear elasticity to
have $\lambda/\mu=1$ and $\nu/\mu=-1$, see Eq.~(12.7) of
\cite{Born88}. Our simulations find $\lambda/\mu=1.21$ and
$\nu/\mu=-1.26$.
\begin{figure}[tb]%
  \centering
  \includegraphics{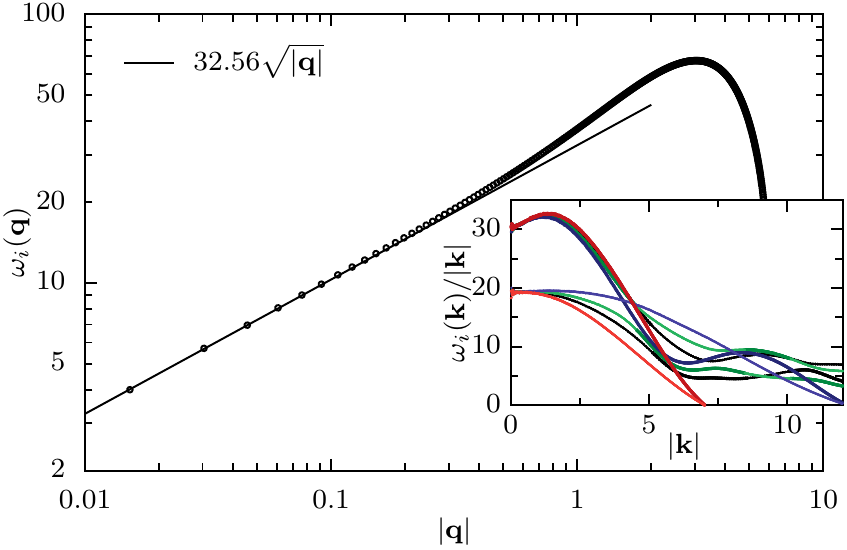}%
  \caption{Inset: full dispersion curves for a two-dimensional
  hexagonal system. The upper curves are longitudinal, the lower ones
  are transverse; directions: red~$(1,0)$, blue~$(2,1)$,
  green~$(3,1)$, black~$(4,1)$. Main plot: Effective dispersion
  relation of a one-dimensional slice in the $(1,0)$ direction,
  compared to the theoretical prediction (solid line). Simulation of
  $400\times400$ particles, $\phi=0.85$.}%
  \label{fig:2D1D}%
\end{figure}%

The use of the Fedorov form for the effective elastic constants
is an uncontrolled approximation. To see to what degree the difference
between theory and simulation is due to this approximation we also
performed calculations and simulations of a two-dimensional ensemble
of disks assembled in a hexagonal lattice projected onto a line. Since
the two-dimensional elastic theory is isotropic~\cite{landau} one
can perform the projection without Fedorov's approximation within linear
elasticity. We measure again the two-dimensional dispersion curves in
the inset of Fig.~\ref{fig:2D1D}, and find a similar effective theory
with $\omega^2 = 4q\mu(\lambda+2\mu)/(\lambda+3\mu)$ for the
projection, which is the longitudinal branch of
Eq.~\eqref{eq:dispersion}. The theoretical value for the coefficient is again
plotted as a line. We find that there is no visible difference
between the theory and the simulations for the projected system.

To conclude, we have shown that if we wish to describe an elastic
slice observed in a confocal microscope as an effective medium we must
introduce an effective dispersion relation in $|\bfq|$ which is very
different from that which occurs in a normal two-dimensional medium
with local interactions. Indeed in real-space one is obliged to
consider that the system has long-ranged effective interactions. These
interactions allow one to avoid Peierls' result implying that a
two-dimensional system should not display long-ranged order, due to
the long-wavelength divergence of the expression for the mean squared
amplitude of positional fluctuations: $\langle u^2 \rangle \sim \int
1/q^2\; d^2\bfq$. This logarithmically diverging result is replaced by
a regular expression due to the change in the dispersion law.

We have shown that to good accuracy one is able to relate the
three-dimensional elastic constants and two-dimensional elastic
behaviour. Thus {\it observations in two dimensions can be used to deduce
estimates of the three-dimensional constants}. It will be particularly
interesting in the future to study how disorder and glassiness modify
these effective properties \cite{kurchan}. It is interesting to note
that such an energy function in $|\bfq|$ was found in \cite{jf} where
the spreading of a droplet was expressed as the effective dynamics of
a contact line. Again we are in the presence of a system projected to
lower dimensions.


\begin{thebibliography}{10}
\expandafter\ifx\csname url\endcsname\relax\def\url#1{\texttt{#1}}\fi

\bibitem{chaikin}
\Name{Cheng Z., Zhu J., Russel W.~B. \and Chaikin P.~M.} \REVIEW{Phys. Rev.
  Lett. }{85}{2000}{1460}.

\bibitem{weeks2}
\Name{Prasad V., Semwogerere D. \and Weeks E.~R.} \REVIEW{Journal of Physics:
  Condensed Matter }{19}{2007}{113102}.
\newline\url{http://stacks.iop.org/0953-8984/19/i=11/a=113102}

\bibitem{maret}
\Name{Zahn K., Wille A., Maret G., Sengupta S. \and Nielaba P.} \REVIEW{Phys.
  Rev. Lett. }{90}{2003}{155506}.

\bibitem{maret2}
\Name{Keim P., Maret G., Herz U. \and von Gr\"unberg H.~H.} \REVIEW{Phys. Rev.
  Lett. }{92}{2004}{215504}.
\newline\url{http://prl.aps.org/abstract/PRL/v92/i21/e215504}

\bibitem{andrea}
\Name{Chen K., Ellenbroek W.~G., Zhang Z., Chen D. T.~N., Yunker P.~J., Henkes
  S., Brito C., Dauchot O., van Saarloos W., Liu A.~J. \and Yodh A.~G.}
  \REVIEW{Phys. Rev. Lett. }{105}{2010}{025501}.

\bibitem{maret3}
\Name{Reinke D., Stark H., von Gr\"unberg H.-H., Schofield A.~B., Maret G. \and
  Gasser U.} \REVIEW{Phys. Rev. Lett. }{98}{2007}{038301}.

\bibitem{antinaPRL}
\Name{Ghosh A., Chikkadi V.~K., Schall P., Kurchan J. \and Bonn D.}
  \REVIEW{Phys. Rev. Lett. }{104}{2010}{248305}.

\bibitem{kurchan}
\Name{Ghosh A., Mari R., Chikkadi V., Schall P., Kurchan J. \and Bonn D.}
  \REVIEW{Soft Matter }{6}{2010}{3082}.
\newline\url{http://dx.doi.org/10.1039/c0sm00265h}

\bibitem{ghosh}
\Name{{Ghosh} A., {Mari} R., {Chikkadi} V.~K., {Schall} P., {Maggs} A.~C. \and
  {Bonn} D.} \REVIEW{Physica~A }{390}{2011}{3061}.
\newline\url{http://www.sciencedirect.com/science/article/pii/S037843711100207%
X}

\bibitem{science}
\Name{Kaya D., Green N.~L., Maloney C.~E. \and Islam M.~F.} \REVIEW{Science
  }{329}{2010}{656}.

\bibitem{danielPRL}
\Name{Ghosh A., Chikkadi V., Schall P. \and Bonn D.} \REVIEW{Phys. Rev. Lett.
  }{107}{2011}{188303}.

\bibitem{ChaLub95}
\Name{Chaikin P.~M. \and Lubensky T.~C.} \Book{Principles of condensed matter
  physics} (Cambridge University Press, Cambridge) 1995 sec.~6.4.

\bibitem{peierls}
\Name{Peierls R.} \Book{Surprises in Theoretical Physics} (Princeton University
  Press, Princeton, N.J.) 1979.

\bibitem{landau}
\Name{Landau L. \and Lifshitz E.} \Book{Theory of Elasticity: Course of
  Theoretical Physics, volume 7, Ch. 1, Section 10.} (Butterworth-Heinemann)
  1984.

\bibitem{Wallace70}
\Name{Wallace D.~C.} \Book{Thermoelastic theory of stressed crystals and
  higher-order elastic constants} in \Book{Solid State Physics. Advances in
  Research and Applications}, edited by \Name{Ehrenreich H., Seitz F. \and
  Turnbull D.} Vol.~25 (Academic Press, New York and London) 1970 pp. 301--404.

\bibitem{Morawiec94}
\Name{Morawiec A.} \REVIEW{Phys.\ Stat.\ Sol~(b) }{184}{1994}{313}.

\bibitem{Fedorov68}
\Name{Fedorov A. F.~I.} \Book{Theory of Elastic Waves in Crystals} (Plenum
  Press, New York) 1968.

\bibitem{Norris06}
\Name{Norris A.~N.} \REVIEW{J.~Acoust.\ Soc.~Am. }{119}{2006}{2114}.

\bibitem{SchMag11}
\Name{Schindler M. \and Maggs A.~C.} \REVIEW{EPJ~E }{34}{2011}{1}.
\newline\url{http://dx.doi.org/10.1140/epje/i2011-11115-7}

\bibitem{rapaport}
\Name{Rapaport D.~C.} \Book{The art of molecular dynamics simulation} 2nd
  Edition (Cambridge Univ. Press) 2004.

\bibitem{frenkel}
\Name{Pronk S. \and Frenkel D.} \REVIEW{Phys. Rev. Lett. }{90}{2003}{255501}.

\bibitem{thesis}
\Name{Ghosh A.} \Book{Thesis} (University of Amsterdam) 2011.

\bibitem{Born88}
\Name{Born M. \and Huang K.} \Book{Dynamical Theory of Crystal Lattices}
  (Oxford University Press, Oxford) 1988.

\bibitem{jf}
\Name{Joanny J.~F. \and de~Gennes P.~G.} \REVIEW{J. Chem. Phys.
  }{81}{1984}{552}.
\newline\url{http://link.aip.org/link/?JCP/81/552/1}

\end{thebibliography}
\end{document}